\documentclass[5p,times,authoryear]{elsarticle}
\usepackage{amsmath,amssymb,amsthm,mathtools,bm}
\usepackage{graphicx,booktabs,array,tabularx}
\usepackage{enumitem,float}
\usepackage[colorlinks=true,allcolors=blue]{hyperref}

\journal{Automatica}
\biboptions{round,authoryear}
\numberwithin{equation}{section}
\newtheorem{theorem}{Theorem}[section]
\newtheorem{lemma}[theorem]{Lemma}
\newtheorem{proposition}[theorem]{Proposition}
\newtheorem{corollary}[theorem]{Corollary}
\theoremstyle{definition}

\theoremstyle{problem}
\newtheorem{problem}[theorem]{Problem}
\theoremstyle{remark}

\setlist[enumerate]{leftmargin=*,itemsep=2pt,topsep=3pt}
\setlist[itemize]{leftmargin=*,itemsep=2pt,topsep=3pt}
\newcommand{\N}{\mathcal N}
\newcommand{\K}{\mathcal K}
\newcommand{\EIn}{E_{UX}}
\newcommand{\EA}{E_{XX}}

\newcommand{\OPT}{\operatorname{OPT}}

\begin{document}
\begin{frontmatter}
\title{Constrained Sparsest Input Selection for Structural Controllability is NP-Hard\tnoteref{funding}}

\author[bit]{Yuan Zhang}
\ead{zhangyuan14@bit.edu.cn}
\author[bit]{Jintao Zhang}
\ead{1120221550@bit.edu.cn}
\author[hunan]{Yun Feng}
\ead{fyrobot@hnu.edu.cn}

\address[bit]{School of Automation, Beijing Institute of Technology, Beijing 100081, China}
\address[hunan]{School of Artificial Intelligence and Robotics, and also National Engineering Research Center for Robot Visual Perception and Control Technology, Hunan University, Changsha 410082, China}
\tnotetext[funding]{This work was supported in part by the National Key Research and Development Program of China under Grant 2025YFA1018800, the National Natural Science Foundation of China under Grant 62373059, and the Beijing Institute of Technology Research Fund Program for Young Scholars.}


\begin{abstract}
We study the problem of selecting the fewest admissible input--state links that make a structured linear system structurally controllable. This objective counts nonzero entries from prescribed positions of the input matrix and is different from minimizing its number of nonzero columns. Prior work gives polynomial-time algorithms for several restricted cases, including single or dedicated inputs and prescribed input patterns satisfying additional graph-theoretic conditions. We give a polynomial-time reduction from perfect three-dimensional matching, establishing for the first time that the decision problem is NP-complete and the optimization problem is NP-hard. A refinement of the reduction yields an exact optimum relation and shows that the problem is APX-hard. An example also exhibits a strict integrality gap for the natural linear programming relaxation.
\end{abstract}
\end{frontmatter}

\section{Introduction}
Structural controllability asks whether a sparsity pattern admits at least one controllable numerical realization and, equivalently, whether almost every admissible realization is controllable. Since Lin's foundational graph-theoretic characterization~\citep{Lin} and its subsequent multi-input developments, the concept has become a standard abstraction for designing actuators/sensors in large-scale systems~\citep{Dion,RamosOverview}. In network applications, the maximum-matching interpretation made it possible to identify minimum sets of driver nodes efficiently~\citep{LiuSlotineBarabasi}. This line of work motivated a broad family of \emph{input-selection} problems, but that phrase now covers several optimization models with materially different decision variables and complexity.

The first model minimizes the number of selected \emph{input channels}, equivalently the number of selected columns from a prescribed collection. General constrained column selection is NP-hard~\citep{PequitoComplexity} and has motivated approximation algorithms~\citep{Moothedath}. Dedicated variants are polynomially solvable by matching methods~\citep{PequitoFramework,Olshevsky,PequitoMinCost}; because every dedicated column contains at most one admissible link, the column and entry objectives coincide in this special case. \citet{ZhangXiaLiuSun} identified broader polynomially solvable column-selection classes through restricted total unimodularity.

The second model, studied here, minimizes the number of selected \emph{input links}, namely
the number of nonzero entries of a structured input matrix $B'$. Every admissible entry may be selected independently.  The distinction is not merely terminological: several selected links can share the same input vertex, and this sharing couples their ability to participate in the bipartite matching required for structural controllability. Thus an NP-hardness result for selecting columns does not establish NP-hardness for selecting entries, nor does a polynomial algorithm for dedicated actuator placement automatically extend to nondedicated link selection. The two objectives coincide only in the dedicated-input case, where each column contains at most one admissible link. 

Polynomial solvability of the entrywise objective is also known in other proper subclasses. \citet{CommaultDion} gave a graph-theoretic polynomial algorithm for the single-input case, minimizing the number of nonzero entries in the input vector. \citet{DeyBalachandranChatterjee} treated prescribed input--state links and obtained polynomial matching- and flow-based algorithms for classes satisfying additional conditions on the state digraph. These results, together with the dedicated-input algorithms cited above, establish tractability for restricted sparsest-link models but do not cover arbitrary nondedicated admissible patterns.
\citet{ZhangZhouPerturbations} studied the broader problem of inserting candidate state and input links, proving NP-hardness and a polynomial-time $2$-approximation. With the state pattern fixed, their approximation bound also applies to the present input-link problem, but their result does not settle its exact complexity.

For entrywise selection, \citet{ZhangXiaZhan} gave integer-programming formulations and proved total unimodularity (TU), exact linear programming (LP) relaxation, and strongly polynomial solvability under the \emph{source-strongly-connected-component grouped input constraint}. However, they also showed that total unimodularity can fail without this condition and left the general link-selection complexity open.

This paper resolves the above-mentioned open complexity question. Its contributions are as follows.
\begin{itemize}
    \item We give an explicit polynomial-time reduction from perfect three-dimensional matching (3DM) to the decision version of constrained sparsest input-link selection. The reduction proves NP-completeness of the decision problem and NP-hardness of the optimization problem for general prescribed input-link patterns.
    \item  Via a refinement of the reduction, we prove that the problem is APX-hard and hence has no polynomial-time approximation scheme unless $\mathrm{P}=\mathrm{NP}$. 
\end{itemize}
Together, these results complement the total-unimodularity result of~\citet{ZhangXiaZhan} by showing that the general constrained link-selection problem is NP-hard.

\section{Preliminaries and Problem Formulation}\label{sec:preliminaries}
\subsection{Structured systems}
Consider the continuous-time linear system
\begin{equation}\label{eq:dynamics}
\dot x(t)=\widetilde A x(t)+\widetilde B u(t),
\end{equation}
where $x(t)\in\mathbb R^n$, $u(t)\in\mathbb R^m$, and $\widetilde A$ and $\widetilde B$ are numerical state and input matrices. A \emph{structured matrix} has entries in $\{0,*\}$: a zero is fixed, whereas $*$ is a free parameter. For a structured matrix $M$, its realization set is $\mathcal S(M)=\{\widetilde M:\widetilde M_{ij}=0\text{ whenever }M_{ij}=0\}$. Let $A\in\{0,*\}^{n\times n}$ and $B\in\{0,*\}^{n\times m}$ specify the sparsity patterns of $\widetilde A$ and $\widetilde B$.

The pair $(A,B)$ is \emph{structurally controllable} if some realization $(\widetilde A,\widetilde B)\in\mathcal S(A)\times\mathcal S(B)$ is controllable. Controllability is generic for this model: whenever one such realization exists, almost every realization is controllable~\citep{Lin,Dion}.

\subsection{Graph terminology and structural controllability}
A directed graph (digraph) is a pair $G=(V,E)$, where $V$ is its vertex set and $E\subseteq V\times V$ its edge set. A directed path from $v_0$ to $v_k$ is a sequence $(v_0,v_1),\ldots,(v_{k-1},v_k)$ of edges. A digraph is strongly connected if every ordered pair of vertices is joined by a directed path. A \emph{strongly connected component} (SCC) is a maximal strongly connected subgraph, and an SCC is a \emph{source SCC} if it has no incoming edge from another SCC.

For $A$, the state digraph is $G(A)=(X,E_A)$, where $X=\{x_1,\ldots,x_n\}$ and $(x_j,x_i)\in E_A$ exactly when $A_{ij}=*$. With input vertices $U=\{u_1,\ldots,u_m\}$, the system digraph is $G(A,B)=(X\cup U,E_A\cup E_B)$, where $(u_j,x_i)\in E_B$ exactly when $B_{ij}=*$. A state vertex is \emph{input-reachable} if a directed path to it starts at some input vertex. All states are input-reachable when every source SCC of $G(A)$ receives an input link.

A bipartite graph has two disjoint vertex classes and edges only between the classes. A \emph{matching} is a set of edges with no common endpoint. The system bipartite graph associated with $(A,B)$ has left class $X_L=\{x_1^L,\ldots,x_n^L\}$ and right class $X_R\cup U$, where $X_R=\{x_1^R,\ldots,x_n^R\}$, and edge set $E_{XX}\cup E_{UX}$, with state edge set $E_{XX}=\{(x_j^R,x_i^L): A_{ij}\ne 0\}$ and input edge set $E_{UX}=\{(u_j,x_i^L): B_{ij}\ne 0\}$. A matching is $X_L$-saturating if every vertex of $X_L$ is matched.

The standard graph criterion~\citep{Dion,RamosOverview} states that $(A,B)$ is structurally controllable if and only if (i) every source SCC of $G(A)$ receives an input link and (ii) the system bipartite graph has an $X_L$-saturating matching. In compact form,
\begin{equation}\label{eq:criterion}
(A,B) \ {\rm \ structurally\ controllable}\iff\text{conditions (i) and (ii) hold}.
\end{equation}

\subsection{Constrained sparsest input-link selection}
For $B$, let $\N(B)=\{(i,j):B_{ij}=*\}$ encode the set of all available input links and $\K(B)=\{B'\in\{0,*\}^{n\times m}:\N(B')\subseteq\N(B)\}$.  Thus $\K(B)$ contains all structured input matrices obtained by retaining an arbitrary subset of the available input links. The optimization problem is
\begin{equation}\label{eq:p1}
\begin{aligned}
(P_1)\quad &\min_{B'\in\K(B)} |\N(B')|,\\
&\text{subject to $(A,B')$ being structurally controllable}.
\end{aligned}
\end{equation}
The objective counts individual nonzero entries: two retained links in one column cost two units, rather than one selected input channel.

\begin{problem}[P1-DEC]\label{def:decision}
Given $(A,B)$ and an integer $K$, decide whether some $B'\in\K(B)$ makes $(A,B')$ structurally controllable and satisfies $|\N(B')|\le K$.
\end{problem}
The objective of this paper is to show that P1-DEC is NP-complete, and therefore $P_1$ is NP-hard. 


\section{Source Problem and Reduction Construction}\label{sec:construction}
\subsection{Perfect three-dimensional matching}
An instance of \emph{perfect three-dimensional matching} consists of three pairwise disjoint sets $\mathcal A=\{a_1,\ldots,a_q\}$, $\mathcal Y=\{b_1,\ldots,b_q\}$, and $\mathcal C=\{c_1,\ldots,c_q\}$, together with triples $T\subseteq\mathcal A\times\mathcal Y\times\mathcal C$.
The question is whether one can select $q$ triples such that no two selected triples agree in any coordinate. Equivalently, a perfect 3DM is a subset $T^\star\subseteq T$ of size $q$ covering every element of each of the three sets exactly once. This decision problem is NP-complete~\citep{GareyJohnson}.

If any element of $\mathcal A$, $\mathcal Y$, or $\mathcal C$ occurs in no triple, the 3DM answer is trivially \emph{no}. Henceforth we assume that every element occurs in at least one triple. This restriction is made only after the stated polynomial-time preprocessing.

\subsection{Construction of the structured system}
Starting from a $T\subseteq \mathcal A\times\mathcal Y\times\mathcal C$, write $N=|T|$. For each $c\in\mathcal C$, list the triples with third coordinate $c$ as $T_c=\{t_{c,1},\ldots,t_{c,k_c}\}$, where $t_{c,j}=(a_{c,j},b_{c,j},c)$ and $k_c\ge1$.
Thus $\sum_{c\in\mathcal C}k_c=N$. Throughout the construction, subscripts of the form $j+1$ are interpreted cyclically modulo $k_c$.

For every triple $t_{c,j}$ introduce a state vertex $v_{c,j}$. For each fixed $c$, arrange these vertices in the directed cycle $v_{c,1}\to v_{c,2}\to\cdots\to v_{c,k_c}\to v_{c,1}$.
Denote this cycle by $C_c$ (edges and vertices in $C_c$ are called cycle edges and cycle states, respectively). When $k_c=1$, it is a single state carrying a self-loop.

For every $b\in\mathcal Y$ create an additional state vertex $w_b$ (called downstream state). For each triple $t_{c,j}=(a,b,c)$ add the state edge $v_{c,j}\to w_b$ (called downstream edge).
No other state edges are included: in particular, no $w_b$ has an outgoing state edge, and no cycle receives a state edge from another component. The number of state vertices is $n=N+q$. Figure~\ref{fig:example} illustrates two cycles and their corresponding downstream edges.


\begin{lemma}[Source SCC structure]\label{lem:scc}
The source SCCs of the constructed state digraph $G(A)$ are exactly the $q$ cycles $\{C_c:c\in\mathcal C\}$.
\end{lemma}
\begin{proof}
Each $C_c$ is strongly connected, because it is a directed cycle. No state edge enters $C_c$ from outside, so it is a source SCC. Distinct cycles have no edges between them. Every $w_b$ is a singleton SCC: it has no outgoing state edge and therefore cannot form a larger SCC with a cycle. Because every $b\in\mathcal Y$ occurs in at least one triple, some downstream edge enters $w_b$ from a cycle. Hence $\{w_b\}$ is not a source SCC. These are all the state vertices, proving the claim.
\end{proof}

For each $a\in\mathcal A$, introduce an input vertex $u_a$. For every triple $t_{c,j}=(a,b,c)$ add the \emph{triple input edge} $u_a\to v_{c,j+1}$. The endpoint is the successor $v_{c,j+1}$, not the state $v_{c,j}$ representing the triple itself. This shift is essential: when the input edge replaces the cycle matching edge $(v_{c,j}^R,v_{c,j+1}^L)$, it frees $v_{c,j}^R$, whose downstream neighbor is exactly the state $w_b$ encoded in the triple.

For each $b\in\mathcal Y$, create a separate \emph{reserve input} vertex $p_b$ and add $p_b\to w_b$.
Reserve inputs will be used only to ensure structural controllability of the \emph{full} system $(A,B)$; they cannot be selected in a solution meeting the reduction's budget. There are $m=2q$ input vertices and $N+q$ allowed input links. Set $K=q$. An illustrative example of such a construction is given in the subsequent subsection as shown in Figure~\ref{fig:example}.

\begin{lemma}[Polynomial construction and feasibility]\label{lem:poly}
The reduction can be performed in polynomial time, and the constructed full-availability pair $(A,B)$ is structurally controllable.
\end{lemma}
\begin{proof}
The construction has $N+q$ state vertices, $2q$ input vertices, $N$ cycle edges, $N$ downstream state edges, $N$ triple input edges, and $q$ reserve input edges. Hence its graph representation has size $O(N+q)$; writing down the corresponding matrices explicitly also takes polynomial time.

For every $c\in\mathcal C$, $T_c$ is nonempty. Consequently at least one triple input edge enters $C_c$. Lemma~\ref{lem:scc} therefore implies that all source SCCs are input-reachable in the full system. To construct a matching saturating all left state vertices, use every cycle edge in every $C_c$. These edges form a matching covering all $N$ cycle state left vertices. In addition, for every $b\in\mathcal Y$ use the reserve input edge $(p_b,w_b^L)$. These $q$ further edges have pairwise distinct left endpoints and input right endpoints, and do not intersect the cycle matching. We obtain a matching of size $N+q=n$. Criterion~\eqref{eq:criterion} establishes that $(A,B)$ is structurally controllable.
\end{proof}

\subsection{An illustration of the construction}
We now give an illustrative example of the construction. 
Let $q=2$. Take $t_1=(a_1,b_1,c_1)$, $t_2=(a_1,b_2,c_2)$, and $t_3=(a_2,b_1,c_2)$, and put $T=\{t_1,t_2,t_3\}$.
This instance has no perfect 3DM. A perfect selection must contain $t_1$ because it is the only triple with third coordinate $c_1$. Selecting $t_2$ as the second triple repeats $a_1$, whereas selecting $t_3$ repeats $b_1$.

Since $c_1$ occurs in only one triple, it produces a singleton source SCC $C_{c_1}=\{v_1\}$ with the self-loop $v_1\to v_1$. The two triples with third coordinate $c_2$ produce the source SCC $C_{c_2}=\{v_2,v_3\}$ with a two-vertex directed cycle $v_2\leftrightarrow v_3$.

\begin{figure}[htbp]
\centering
\includegraphics[width=1\linewidth]{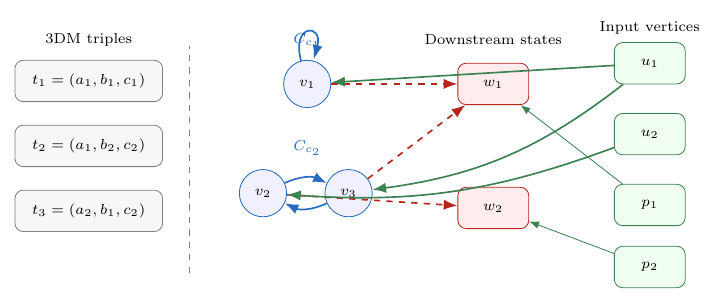}
\caption{A 3DM-to-structural-controllability reduction. The source SCCs encode $c$, input vertices $u_a$ encode $a$, and downstream states $w_b$ encode $b$.}
\label{fig:example}
\end{figure}

The downstream state vertices are $w_1,w_2$, with edges $v_1\to w_1$, $v_2\to w_2$, and $v_3\to w_1$. The triple input edges are $u_1\to v_1$, $u_1\to v_3$, and $u_2\to v_2$.
Finally, introduce reserve links $p_1\to w_1$ and $p_2\to w_2$. Figure~\ref{fig:example} illustrates all these edges. Solid blue edges are internal cycle edges, dashed red edges are downstream state edges, and green edges are input links (including the reserve links).

\noindent Taking the state order $(v_1,v_2,v_3,w_1,w_2)$ and the input order $(u_1,u_2,p_1,p_2)$, we obtain the concrete structured matrices
\begin{equation}\label{eq:exampleAB}
A=\begin{bmatrix}
 *&0&0&0&0\\
 0&0&*&0&0\\
 0&*&0&0&0\\
 *&0&*&0&0\\
 0&*&0&0&0
\end{bmatrix},\qquad
B=\begin{bmatrix}
 *&0&0&0\\
 0&*&0&0\\
 *&0&0&0\\
 0&0&*&0\\
 0&0&0&*
\end{bmatrix}.
\end{equation}
The entries of $A$ are determined by the convention $A_{ij}=*$ when the state edge is $x_j\to x_i$. As shown subsequently, the fact that this instance has no perfect 3DM implies that every structurally controllable input-link selection $S$ for $(A,B)$ satisfies $|S|>q$.

\section{Correctness of the Reduction}\label{sec:correctness}
\subsection{Perfect 3DM implies budget feasibility}
\begin{lemma}\label{lem:forward}
If the 3DM instance admits a perfect three-dimensional matching, then the constructed P1-DEC instance admits a structurally controllable selection of exactly $q$ input links.
\end{lemma}
\begin{proof}
Let $T^\star\subseteq T$ be a perfect 3DM. Denote its selected triple for $c$ by $t_{c,j_c}=(a_c,b_c,c)$, and select the corresponding input links $S^\star=\{(u_{a_c},v_{c,j_c+1}):c\in\mathcal C\}$.
Each cycle $C_c$ receives one selected input edge, so source SCC reachability holds. We construct a matching saturating all left state vertices using only the input links in $S^\star$.

Fix $c$. Include $(u_{a_c},v_{c,j_c+1}^L)$ and all $k_c-1$ cycle matching edges except $(v_{c,j_c}^R,v_{c,j_c+1}^L)$. These edges cover every left state vertex of $C_c$ and leave $v_{c,j_c}^R$ unused. Include $(v_{c,j_c}^R,w_{b_c}^L)$, which exists because $t_{c,j_c}=(a_c,b_c,c)$.

The constructions for different values of $c$ are compatible. cycle state vertices from different cycles are disjoint; the input vertices $u_{a_c}$ are distinct because the first coordinates are distinct; and the downstream vertices $w_{b_c}^L$ are distinct because the second coordinates are distinct. The resulting matching saturates $\sum_c k_c+q=N+q=n$ left state vertices. By~\eqref{eq:criterion}, the selected system is structurally controllable and uses exactly $q=K$ input links. If $k_c=1$, the same argument applies: the input edge replaces the single self-loop matching edge, freeing its state right vertex to match the associated downstream state.
\end{proof}

\subsection{Budget feasibility implies perfect 3DM}
\begin{lemma}\label{lem:reverse}
If the constructed P1-DEC instance has a structurally controllable input-link selection $S$ with $|S|\le q$, then the original 3DM instance has a perfect three-dimensional matching.
\end{lemma}
\begin{proof}
The proof proceeds in four steps.

\smallskip\noindent\textbf{Step 1: Exactly one link enters each source SCC.}
By Lemma~\ref{lem:scc}, the state graph has exactly $q$ source SCCs. Each requires a selected input link, and one link can enter only one source SCC, so $|S|\ge q$. Together with $|S|\le q$, this gives $|S|=q$ and $|S\cap E(C_c)|=1$ for every $c\in\mathcal C$, where $E(C_c)$ denotes the allowed input links entering $C_c$. No reserve link is selected because it enters no source SCC.

\smallskip\noindent\textbf{Step 2: All $q$ selected input links belong to the saturating matching.}
Let $M\subseteq E_{XX}\cup E_{UX}$ saturate all $n=N+q$ left state vertices. Since the downstream states have no outgoing state edges, only the $N$ cycle state right vertices support state matching edges. Thus $|M\cap\EA|\le N$ and $|M\cap\EIn|\ge q$. Every input matching edge must be selected and $|S|=q$, so
\begin{equation}\label{eq:allinput}
|M\cap\EIn|=q,\qquad
M\cap\EIn=S,\qquad
|M\cap\EA|=N.
\end{equation}
Since a matching cannot reuse an input right vertex, the selected triple input edges have pairwise distinct input vertices $u_a$. Moreover, all $N$ cycle state right vertices are used by state matching edges in $M$.

\smallskip\noindent\textbf{Step 3: Each selected input link uniquely releases the predecessor right vertex.}
Fix $c$. The unique link entering $C_c$ is $(u_{a_{c,j}},v_{c,j+1}^L)$ for the triple $t_{c,j}=(a_{c,j},b_{c,j},c)$, and~\eqref{eq:allinput} puts it in $M$. Every other cycle state left vertex in $C_c$ has a unique state-matching predecessor. Thus $M$ uses all $k_c-1$ cycle edges except $(v_{c,j}^R,v_{c,j+1}^L)$.
The unique cycle state right vertex not used by these internal edges is $v_{c,j}^R$.

By~\eqref{eq:allinput}, all $N$ cycle state right vertices must be used by state matching edges. Hence $v_{c,j}^R$ is forced to use its only edge leaving the cycle, $(v_{c,j}^R,w_{b_{c,j}}^L)$, corresponding to $t_{c,j}$. This also covers $k_c=1$.

\smallskip\noindent\textbf{Step 4: Recover a perfect three-dimensional matching.}
For each $c\in\mathcal C$, let $t_c=(a_c,b_c,c)$ be the triple corresponding to the unique selected input edge entering $C_c$, and set $T^\star=\{t_c:c\in\mathcal C\}$.
These $q$ triples have distinct third coordinates by construction. If two have the same first coordinate $a$, their corresponding selected input edges share the input vertex $u_a$. Both belong to $M$ by~\eqref{eq:allinput}, contradicting the matching property. Thus their first coordinates are distinct.

If two selected triples have the same second coordinate $b$, Step 3 forces two different state matching edges to end at the same downstream left vertex $w_b^L$. Again this contradicts the matching property. Thus their second coordinates are distinct. Since each of the three coordinate sets contains $q$ elements, $T^\star$ is a perfect three-dimensional matching.
\end{proof}

\section{NP-Completeness and Complexity Consequences}
\begin{theorem}\label{thm:main}
The decision problem P1-DEC is NP-complete for general prescribed input-link patterns. Consequently, the optimization problem $P_1$ is NP-hard.
\end{theorem}
\begin{proof}
Given a 3DM instance, construct $(A,B,K)$ as in Section~\ref{sec:construction} with $K=q$. The construction is polynomial-time by Lemma~\ref{lem:poly}. Lemmas~\ref{lem:forward} and~\ref{lem:reverse} yield
\begin{equation}\label{eq:equiv}
T\text{ has a perfect 3DM}
\quad\Longleftrightarrow\quad
\OPT(P_1)\le q.
\end{equation}
Hence $\mathrm{3DM}\le_p\mathrm{P1\text{-}DEC}$ and P1-DEC is NP-hard.

To show membership in NP, use the positions of the $*$ entries of $B'$ as a polynomial-size certificate. Checking $B'\in\K(B)$ and $|\N(B')|\le K$ is polynomial. Structural controllability can also be verified in polynomial time~\citep{RamosOverview}. Thus P1-DEC belongs to NP and is NP-complete.
\end{proof}



For completeness, we recall the integer linear programming (ILP) of $P_1$ from~\citet{ZhangXiaZhan}. Let $E=\EA\cup\EIn$, let $\mathcal I$ index the source SCCs of $G(A)$, let $r=|\mathcal I|$, and let $E_i\subseteq\EIn$ be the input links entering source SCC $i\in\mathcal I$. Introduce binary variables $y_{uv}$ for $(u,v)\in E$, where $y_{uv}=1$ means that $(u,v)$ belongs to the selected saturating matching, and binary variables $z_i$, where $z_i=1$ means that this matching contains an input link in $E_i$. The ILP formulation is
\begin{equation}\label{eq:p1ilp}
\begin{aligned}
\min_{y,z}\quad
 &\sum_{(u,v)\in\EIn}y_{uv}+r-\sum_{i\in\mathcal I}z_i \\
\mathrm{s.t.}\quad
 &\sum_{u:(u,v)\in E}y_{uv}=1,
 &&\forall v\in X_L,\\
 &\sum_{v:(u,v)\in E}y_{uv}\le1,
 &&\forall u\in X_R\cup U,\\
 &z_i\le\sum_{(u,v)\in E_i}y_{uv},
 &&\forall i\in\mathcal I,\\
 &y_{uv}\in\{0,1\},\quad z_i\in\{0,1\}.
\end{aligned}
\end{equation}
From \citet{ZhangXiaZhan}, the optimal value of problem \eqref{eq:p1ilp} equals $\OPT(P_1)$.  The natural LP relaxation replaces the binary restrictions in~\eqref{eq:p1ilp} by $0\le y_{uv}\le1$ and $0\le z_i\le1$.
We now apply this relaxation to the five-state instance in Figure~\ref{fig:example}, for which $\mathcal I=\{c_1,c_2\}$ and $r=2$. Let $y_e=1/2$ for the following ten edges:
\begin{align*}
 &(v_1^R,v_1^L),\ (v_2^R,v_3^L),(v_3^R,v_2^L),
 (v_1^R,w_1^L),\ (v_2^R,w_2^L),\ (v_3^R,w_1^L),\\
 &(u_1,v_1^L),\ (u_1,v_3^L),\ (u_2,v_2^L),\ (p_2,w_2^L),
\end{align*}
set all other $y_e=0$, and take $z_{c_1}=1/2$ and $z_{c_2}=1$.
Every left state vertex has total incident matching weight exactly one, and every right vertex has total incident matching weight at most one. The sum of input matching weights entering $C_{c_1}$ is $1/2$, and that entering $C_{c_2}$ is $1$, so the two $z$ inequalities hold. Exactly four input edges have weight $1/2$, yielding a total input matching weight of $2$. Therefore the relaxed objective is
\begin{equation}\label{eq:gap}
\sum_{e\in\EIn}y_e+r-\sum_{c\in\mathcal C}z_c
=2+2-\left(\frac12+1\right)=\frac52
<3=\OPT(P_1).
\end{equation}
Consequently, the LP relaxation has a strict integrality gap on this instance. This implies that the LP-relaxation-based method in \citet{ZhangXiaZhan} may fail outside the so-called source SCC grouped input constraint there (each input node cannot be linked to more than one source SCC).  

\section{Approximability}\label{sec:approximation}
For a minimization problem, a polynomial-time algorithm is a $\rho$-approximation ($\rho\ge1$) if every returned feasible value $f$ satisfies
\begin{equation}\label{eq:approxdef}
\OPT\le f\le\rho\OPT.
\end{equation}
A \emph{polynomial-time approximation scheme} (PTAS) gives a $(1+\varepsilon)$-approximation for every fixed $\varepsilon>0$, in time polynomial in the input size. The class APX contains problems admitting a constant-factor approximation. A problem is \emph{APX-hard} if every problem in APX can be transferred to it by an approximation-preserving reduction (L-reduction).\footnote{An \emph{L-reduction} from an optimization problem \(A\) to an optimization problem \(B\)
consists of two polynomial-time computable functions \(f\) and \(g\),
together with constants \(\alpha,\beta>0\), such that for every instance \(x\) of \(A\), \(f(x)\) is an instance of \(B\) satisfying
  $
    \mathrm{OPT}_B(f(x)) \le \alpha \cdot \mathrm{OPT}_A(x)
  $, and for every feasible solution \(y\) of \(f(x)\),
  $
    \left|\mathrm{OPT}_A(x) - c_A(g(x,y))\right|
    \le
    \beta \cdot \left|\mathrm{OPT}_B(f(x)) - c_B(y)\right|,
  $ where \(c_A\) and \(c_B\) are the cost functions of \(A\) and \(B\),
  and \(g(x,y)\) is a feasible solution of \(x\).} Unless $\mathrm{P}=\mathrm{NP}$, an APX-hard problem has no PTAS; this is compatible with the existence of a fixed-factor algorithm, such as the 2-approximation discussed in \citet{ZhangZhouPerturbations}.

The present reduction preserves more than the yes--no threshold. Let $\nu(T)$ be the maximum cardinality of a three-dimensional matching contained in $T$.

\begin{proposition}[Exact optimum relation]\label{prop:approx}
For the instance constructed from $T$ in Section \ref{sec:construction},
\begin{equation}\label{eq:exactopt}
\OPT(P_1)=2q-\nu(T).
\end{equation}
Moreover, from any feasible input-link selection of cardinality $s$ one can recover in polynomial time a three-dimensional matching of cardinality at least $2q-s$.
\end{proposition}
\begin{proof}
Let $T_M$ be a three-dimensional matching of size $\nu(T)$. Select its $\nu(T)$ triple input links, an arbitrary triple input link for each of the $q-\nu(T)$ uncovered third coordinates, and a reserve link for each of the $q-\nu(T)$ uncovered second coordinates. The total is $2q-\nu(T)$. Every source SCC is reached, and a saturating matching follows by using the replacement construction of Lemma~\ref{lem:forward} for $T_M$, the complete cycle matching elsewhere, and the selected reserve links. Thus $\OPT(P_1)\le2q-\nu(T)$.

Conversely, let $S$ be feasible with $|S|=s$, and let $r$ selected links be reserve links. Reachability supplies a set $Q$ of $q$ selected triple links, one entering each source SCC. In an $X_L$-saturating matching, at most $r$ downstream vertices can be matched by reserve inputs. Each of the remaining at least $q-r$ downstream matching edges $(v_{c,j}^R,w_b^L)$ forces the corresponding selected triple input edge $(u_a,v_{c,j+1}^L)$. Let $P$ collect these links. Their first coordinates are distinct because their input endpoints are matched, their second coordinates are distinct because their downstream endpoints are matched, and $|P|\ge q-r$. Since there are $s-r$ selected triple links,
\begin{equation}\label{eq:recoverbound}
|P\cap Q|\ge |P|+|Q|-(s-r)\ge2q-s.
\end{equation}
The triples represented by $P\cap Q$ also have distinct third coordinates because they lie in $Q$. Hence they form a 3DM of size at least $2q-s$, implying $s\ge2q-\nu(T)$ and proving~\eqref{eq:exactopt}.
\end{proof}

\begin{corollary}[APX-hardness]\label{cor:apx}
The constrained sparsest input-link selection problem is APX-hard, even when the full available system is structurally controllable. Consequently, it has no PTAS unless $\mathrm{P}=\mathrm{NP}$.
\end{corollary}
\begin{proof}
Maximum bounded-occurrence 3DM is  APX-hard~\citep{Kann}. On its standard hard family, every element occurs in at most a fixed constant $C$ and may be assumed to occur at least once. A maximal 3DM (maximal means no additional elements from $T$ can be added to this 3DM to increase its size) then has size at least $q/(3C)$, so $q\le3C\nu(T)$. Proposition~\ref{prop:approx} gives $\OPT(P_1)=2q-\nu(T)\le6C\nu(T)$. From any feasible target solution of cost $s$, the recovered matching has size $\mu\ge2q-s$, whence $\nu(T)-\mu\le s-\OPT(P_1)$. The construction and recovery map are thus an L-reduction \citep{Kann} with constants depending only on $C$. Lemma~\ref{lem:poly} preserves full-system feasibility.
\end{proof}

\section{Conclusions and Scope}

We prove that the constrained sparsest input-link selection problem is NP-hard. This complements the polynomial algorithms of~\citet{ZhangXiaZhan} and~\citet{PequitoFramework}, which rely on the source SCC grouped input constraint or dedicated input structure, by establishing NP-hardness when no such restriction is imposed. The exact relation~\eqref{eq:exactopt} further transfers the APX-hardness of bounded-occurrence maximum 3DM, ruling out a PTAS unless $\mathrm{P}=\mathrm{NP}$. These conclusions do not contradict the tractable dedicated, restricted-TU, or grouped-input cases; rather, they identify a genuine complexity boundary between those structured subclasses and general nondedicated link selection.



\begin{thebibliography}{99}
{\small{
\bibitem[Commault and Dion(2015)]{CommaultDion}
C. Commault and J.-M. Dion,
``The single-input minimal controllability problem for structured systems,''
\emph{Systems \& Control Letters}, vol.~80, pp.~50--55, 2015.

\bibitem[Dey et~al.(2018)Dey, Balachandran, and Chatterjee]{DeyBalachandranChatterjee}
P. Dey, N. Balachandran, and D. Chatterjee,
``On minimum cost sparsest input-connectivity for controllability of linear systems,''
in \emph{Proceedings of the 57th Annual Conference of the Society of Instrument and Control Engineers of Japan (SICE)}, 2018, doi:10.23919/SICE.2018.8492603.

\bibitem[Dion et~al.(2003)Dion, Commault, and van der Woude]{Dion}
J.-M. Dion, C. Commault, and J. van der Woude,
``Generic properties and control of linear structured systems: A survey,''
\emph{Automatica}, vol. 39, pp. 1125--1144, 2003.

\bibitem[Garey and Johnson(1979)]{GareyJohnson}
M. R. Garey and D. S. Johnson,
\emph{Computers and Intractability: A Guide to the Theory of NP-Completeness},
W. H. Freeman, 1979.

\bibitem[Kann(1991)]{Kann}
V. Kann,
``Maximum bounded 3-dimensional matching is MAX SNP-complete,''
\emph{Information Processing Letters}, vol.~37, no.~1, pp.~27--35, 1991.

\bibitem[Lin(1974)]{Lin}
C.-T. Lin,
``Structural controllability,''
\emph{IEEE Transactions on Automatic Control}, vol.~19, no.~3, pp.~201--208, 1974.

\bibitem[Liu et~al.(2011)Liu, Slotine, and Barab\'{a}si]{LiuSlotineBarabasi}
Y.-Y. Liu, J.-J. Slotine, and A.-L. Barab\'{a}si,
``Controllability of complex networks,''
\emph{Nature}, vol.~473, pp.~167--173, 2011.

\bibitem[Moothedath et~al.(2018)Moothedath, Chaporkar, and Belur]{Moothedath}
S. Moothedath, P. Chaporkar, and M. N. Belur,
``A flow-network-based polynomial-time approximation algorithm for the minimum constrained input structural controllability problem,''
\emph{IEEE Transactions on Automatic Control}, vol.~63, no.~9, pp.~3151--3158, 2018.

\bibitem[Olshevsky(2015)]{Olshevsky}
A. Olshevsky,
``Minimum input selection for structural controllability,''
in \emph{Proceedings of the 2015 American Control Conference}, pp.~2218--2223, 2015.

\bibitem[Pequito et~al.(2015)Pequito, Kar, and Aguiar]{PequitoComplexity}
S. Pequito, S. Kar, and A. P. Aguiar,
``On the complexity of the constrained input selection problem for structural linear systems,''
\emph{Automatica}, vol.~62, pp.~193--199, 2015.

\bibitem[Pequito et~al.(2016a)Pequito, Kar, and Aguiar]{PequitoFramework}
S. Pequito, S. Kar, and A. P. Aguiar,
``A framework for structural input/output and control configuration selection in large-scale systems,''
\emph{IEEE Transactions on Automatic Control}, vol.~61, no.~2, pp.~303--318, 2016.

\bibitem[Pequito et~al.(2016b)Pequito, Kar, and Aguiar]{PequitoMinCost}
S. Pequito, S. Kar, and A. P. Aguiar,
``Minimum cost input/output design for large-scale linear structural systems,''
\emph{Automatica}, vol.~68, pp.~384--391, 2016.

\bibitem[Ramos et~al.(2022)Ramos, Aguiar, and Pequito]{RamosOverview}
G. Ramos, A. P. Aguiar, and S. Pequito,
``An overview of structural systems theory,''
\emph{Automatica}, vol.~140, Art.~no.~110229, 2022.

\bibitem[Zhang et~al.(2023)Zhang, Xia, Liu, and Sun]{ZhangXiaLiuSun}
Y. Zhang, Y. Xia, S. Liu, and Z. Sun,
``On polynomially solvable constrained input selections for fixed and switched linear structured systems,''
\emph{Automatica}, vol.~158, Art.~no.~111308, 2023.

\bibitem[Zhang et~al.(2024)Zhang, Xia, and Zhan]{ZhangXiaZhan}
Y. Zhang, Y. Xia, and Y. Zhan,
``Total unimodularity and strongly polynomial solvability of constrained minimum input selections for structural controllability: An LP-based method,''
\emph{IEEE Transactions on Automatic Control}, vol.~69, no.~1, pp.~387--394, 2024.

\bibitem[Zhang and Zhou(2019)]{ZhangZhouPerturbations}
Y. Zhang and T. Zhou,
``Minimal structural perturbations for controllability of a networked system: Complexities and approximations,''
\emph{International Journal of Robust and Nonlinear Control}, vol.~29, no.~12, pp.~4191--4208, 2019.}}
\end{thebibliography}
\end{document}